\documentstyle[twocolumn,eqsecnum,aps]{revtex}
\def\be{\begin{equation}}
\def\ee{\end{equation}}
\def\bea{\begin{eqnarray}}
\def\eea{\end{eqnarray}}
\begin{document}
\draft
\title{Impact Parameter Dependent Parton Distributions and
Off-Forward Parton Distributions for $\zeta\rightarrow 0$}
\author{Matthias Burkardt}
\address{Department of Physics\\
New Mexico State University\\
Las Cruces, NM 88003-0001\\U.S.A.}
\maketitle
\begin{abstract}
It is shown that off-forward parton distributions for $\zeta =0$, 
i.e. where the initial and final state differ only in their
transverse momenta, can be interpreted in terms of a simultaneous 
measurement of the longitudinal momentum and transverse position 
(impact parameter) of partons in the infinite momentum frame.
\end{abstract}
\narrowtext
\section{Introduction}
Deeply virtual Compton scattering (DVCS) provides a novel tool 
to explore hadron structure. In contrast to deep-inelastic 
scattering (DIS), where one measures the imaginary part of the 
forward Compton amplitude only, DVCS allows measuring the 
off-forward Compton amplitude. From the parton point of view 
this implies that DVCS allows measuring off-forward matrix 
elements of parton correlation functions, i.e. on can access 
light-cone correlation functions of the form \cite{ji,ar}.
\footnote{For some history on DVCS see Ref. \cite{wally}.}
\be
f_\zeta(x,t) \equiv \int \frac{dx^-}{4\pi} 
\langle p^\prime | \bar{\psi}(0) \gamma^+
\psi(x^-) |p\rangle e^{ixp^+x^-} , \label{eq:off}
\ee
where $x^\pm = x^0\pm x^3$ and $p^+=p^0+p^3$ refers to the usual 
light-cone components and $t\equiv q^2=(p-p^\prime )^2$ 
is the invariant momentum transfer. The ``off-forwardness'' 
(or skewedness) in Eq. (\ref{eq:off}) is defined to be 
$\zeta\equiv \frac{q^+}{p^+}$.
From the point of view of parton physics in the infinite momentum
frame, these off-forward parton distributions (OFPDs) have the
physical meaning of the amplitude for the process that a quark 
is taken out of the nucleon with momentum fraction $x$ and then 
it is inserted back into the nucleon with a four momentum 
transfer $q^\mu$ \cite{wally}.
It was immediately recognized that off-forward parton distribution play
a dual roles in that they combine features of both form factors and
conventional parton distribution functions \cite{ji,ar}:
for $\zeta=t=0$ one recovers conventional parton distributions, i.e.
momentum distributions in the infinite momentum frame (IMF), while
when one integrates $f_\zeta(x,t)$ over $x$, one obtains a form factor,
i.e. the Fourier transform of a coordinate space density (in the Breit frame!).

However, the physical interpretation of the general case still remained
obscure, mainly because the initial and final state in
Eq. (\ref{eq:off}) are not the same and therefore, in general, 
$f_\zeta(x,t)$ cannot be interpreted as a `density' but
rather their physical significance is that of a 
probability amplitude.

In this note, we will study a more general limiting case, namely 
$\zeta=0$, but $t\neq 0$, and we will argue that 
\be
f(x,t)\equiv f_{\zeta=0}(x,t)
\ee
has a simple interpretation in terms of a density as well, namely
as the Fourier transform of the light-cone momentum/impact parameter
density w.r.t. the impact parameter.

In a light-front (LF) framework it is easy to see that the case
$\zeta=0$ is particularly simple since there only terms diagonal 
in the Fock space contribute to $f(x,t)$ (just as it is the case 
for ordinary parton distributions). Explicit Fock space representations 
for $f(x,t)$ can be found in Refs. \cite{kroll1,kroll2}. 
Making a Gaussian ansatz for the Fock space components, it is therefore 
straightforward to see the connection between the $t$-dependence of $f(x,t)$ 
and the Gaussian size parameter \cite{kroll2,ar3}.
In this paper, we will demonstrate that this very intuitive connection
is valid independent of specific models and that it is in fact possible
to determine parton distributions as a function of the impact parameter,
provided $f(x,t)$ is known.

Of course, $\zeta=0$ with $t\neq 0$ cannot be achieved in
virtual Compton scattering at finite energies because it
always takes some longitudinal momentum transfer in order
to convert a virtual photon into a real photon, i.e.
strictly speaking $\zeta=0$ would correspond to real (wide angle)
Compton scattering \cite{ar3,ar2}. However, as a limiting case 
($\zeta \rightarrow 0$), $f(x,t)$ is relevant for 
DVCS as well.

The paper is organized as follows. In Section II, we
use a familiar observable, the elastic charge form factor,
to illustrate how relativistic effects may spoil the
identification of Fourier transforms of position space 
distributions with form factors. Since this is a well
known phenomenon for form factors, Section II will mainly
serve to introduce our notation and reasoning.
In Section III, we will then generalize the results from
Section II to $\zeta=0$ OFPDs. Section III contains the 
derivation of the main result of this paper, namely the
identification of Fourier transforms of impact parameter
dependent parton distributions with OFPDs. The results are
summarized in Section IV.

\section{Form Factors and Charge Distributions}
Nonrelativistic intuition suggests to interpret ordinary
charge form factors as Fourier transforms of charge 
distributions in position space. As a warmup exercise, we will
in the following reexamine the limitations of this interpretation
in a relativistic framework.

Since the charge distribution of a plane wave is ill defined,
it is useful to start from a wave packet \footnote{I would like to thank 
Bob Jaffe for his suggestion to use wave packets in the discussion
of relativistic corrections.}
\be
\left| \Psi \right\rangle = \int \frac{d^3p\,\,\,
\psi({\vec p}) }{\sqrt{2 E_{\vec p} 
(2\pi)^3}} \left| {\vec p}\right\rangle ,
\ee
where $E_{\vec p}=\sqrt{M^2+{\vec p}^2}$.
Momentum space eigenstates are normalized covariantly as usual, 
i.e. $\langle {\vec p}^\prime |{\vec p} \rangle = 2E_{\vec p}
\delta({\vec p}^\prime -{\vec p} )$.
Using the usual definition of the charge form factor
\be
\left\langle {\vec p}^\prime \right|
\rho({\vec 0})\left|{\vec p}\right\rangle
= \left(E_{\vec p}+E_{{\vec p}^\prime}\right)
F(q^2),
\ee
where 
\be
q^2=( E_{\vec p}-E_{{\vec p}^\prime})^2-
{\vec q}^2 
\label{eq:q2}
\ee
and ${\vec q}= {\vec p}^\prime -{\vec p}$,
one obtains for the Fourier transform of the charge
distribution in the wave packet
\bea
{\cal F}_{\psi}({\vec q})&\equiv& \int d^3x e^{-i{\vec q}
\cdot {\vec x}} \left\langle \Psi \right| \rho({\vec x})
\left| \Psi \right\rangle
\nonumber\\
&=& \int \frac{d^3p}{\sqrt{2E_{\vec p} 2E_{{\vec p}^\prime}}}
\Psi^*({\vec p}+{\vec q}) \Psi({\vec p})
\left\langle {\vec p}^\prime \right|
\rho({\vec 0})\left|{\vec p}\right\rangle \nonumber\\
&=& \frac{1}{2} \int d^3p \frac{ E_{\vec p}+E_{{\vec p}^\prime} }
{\sqrt{ E_{\vec p} E_{{\vec p}^\prime}}}
\Psi^*({\vec p}+{\vec q}) \Psi({\vec p}) F(q^2) .
\label{eq:con}
\eea
Note that $q^2$ still depends implicitly on ${\vec p}$
(\ref{eq:q2}) and thus in general one cannot pull
$F(q^2)$ out of the integral! Eq. (\ref{eq:con}) clearly
illustrates how the charge distribution in the wave packet
is obtained from a convolution of the Form factor with the
spatial distribution of the wave packet as well as various
relativistic effects. 

Initially, the wave packet was only introduced in order to
be able to cleanly define a charge distribution. On the other
hand, we are interested only in the intrinsic charge distribution
of the hadron, i.e. not in the distribution due to the wave
packet and therefore we would like to get rid of anything
associated with the wave packet in Eq. (\ref{eq:con}).

\subsection{Nonrelativistic Limit}
In the nonrelativistic limit, where 
$\frac{ E_{\vec p}+E_{{\vec p}^\prime} }
{2\sqrt{ E_{\vec p} E_{{\vec p}^\prime}}}
=1$ and $q^2=-{\vec q}^2$ one can pull the form factor
out of the integral in Eq. (\ref{eq:con}), yielding
\be
{\cal F}_{\psi}({\vec q})
= F(-{\vec q}^2) \int d^3p 
\Psi^*({\vec p}+{\vec q}) \Psi({\vec p}) .
\label{eq:connr}
\ee
Finally, by using a wave packet $\Psi$ that is very broad
in momentum space (i.e. localized in position space!)
the dependence of the overlap integral $\int d^3p 
\Psi^*({\vec p}+{\vec q}) \Psi({\vec p})$ on ${\vec q}$ 
is much weaker than the dependence of the form factor 
$F(-{\vec q}^2)$. 
For such a wave packet Eq. (\ref{eq:connr})
one thus finds 
\be
{\cal F}_\Psi ({\vec q}) = F(-{\vec q}^2) \int d^3p 
\left|\Psi({\vec p})\right|^2= F({\vec q}^2)
\label{eq:formnr}
\ee
Therefore, in a nonrelativistic theory,
\footnote{Actually, `nonrelativistic' is necessary here only to the extent
that the momentum transfer leaves the target nonrelativistic!}
as long as one uses 
a wave packet which is very localized in
position space, the Fourier transform of the 
charge distribution for this wave packet equals the form factor.
It is thus legitimate to interpret the
form factor as the Fourier transform of the intrinsic charge
distribution.

\subsection{Relativistic Corrections (Rest Frame)}
Unfortunately, in a relativistic theory, it is in general
not possible to form a wave packet of states
whose Fourier transformed
charge distribution equals the form factor.
In the nonrelativistic case, we used a wave packet that had
an arbitrarily small extension in position space.
In a relativistic theory, localizing a wave packet to less than 
its Compton 
wavelength in size will in general induce various relativistic
corrections. This fact is best illustrated by considering 
the rms radius of this charge distribution
[$2^{nd}$ derivative w.r.t. ${\vec q}$ in Eq. (\ref{eq:con})]

Expanding ${\cal F}_\Psi({\vec q})$ up to, and including
${\cal O}({\vec q}^2)$, one finds \footnote{We assume
here a target with unit charge. The generalization to
other values for the charge is straightforward.}
\bea
{\cal F}_\Psi({\vec q})&=& 1 - \frac{R^2}{6}{\vec q}^2 
-\frac{R^2}{6}\int d^3p \left|\Psi({\vec p})\right|^2
\frac{ \left({\vec q}\cdot {\vec p} \right)^2}
{E^2_{\vec p}} 
\label{eq:rms}\\
&+& 1 \int d^3p 
\left| {\vec q}\cdot{\vec \nabla}\Psi({\vec p}) \right|^2
- \frac{1}{8} \int d^3p \left|\Psi({\vec p})\right|^2
\frac{\left({\vec q}\cdot {\vec p} \right)^2}
{E_{\vec p}^4},
\nonumber
\eea
where $R^2$ is defined through the slope of $F(q^2)=1+\frac{R^2}{6}q^2
+{\cal O}(q^4)$.
In addition to the contribution from the intrinsic
size and the contribution from the size of the wave packet,
one obtains a Lorentz contraction contribution and other
relativistic corrections.

Ideally, one would again like to construct a wave packet such 
that the contribution from the spatial extension of the
wave packet, i.e. the term $\propto \int d^3p 
\left| {\vec q}\cdot{\vec \nabla}\Psi({\vec p}) \right|^2$,
is negligible compared to $R^2{\vec q}^2$.
Making the corrections due to the extension of the wave packet
negligible requires a typical
momentum scale in $\Psi({\vec p})$ that is much larger than
$\frac{1}{R}$. This on the other hand leads to 
contributions from the relativistic corrections in Eq.
(\ref{eq:rms}) that are at least of the order
$\Delta R^2 \sim \frac{1}{R^2M^4}$, which are negligible only
if the Compton wavelength of the target is much smaller
than its intrinsic size (as defined through the slope of the
form factor).

The physics of this result is clear: as soon as one attempts
to localize the wave packet to a region smaller than its
Compton wavelength, the particle in the wave packet becomes
relativistic and relativistic effects, such as Lorentz 
contraction, are no longer negligible.
What this means is that an identification of the
slope of the formfactor with the rms-radius of a charge
distribution in the rest frame is not unambiguously possible
and the best one can achieve is an identification with
uncertainties on the order of the Compton wavelength of the 
target.

\subsection{Infinite Momentum Frame}
In certain frames, such as the IMF
(which will be relevant for the application to
off-forward parton distribution!), this ambiguity can
be avoided. The essential point is that the relativistic
corrections are governed by coefficients like
$\frac{{\vec q}\cdot {\vec p}}{E_{\vec p}^2}$
and $\frac{{\vec q}^2}{E_{\vec p}^2}$.
One way to keep these relativistic coefficients small is 
to keep ${\vec q}^2$ and
${\vec q}\cdot {\vec p} $ finite while sending
$ E_{\vec p}\rightarrow \infty$, i.e. by going to the
IMF!
In the following, let us assume a wave packet such that 
${\vec P}=(0,0,p_z)$ is the mean momentum of the wave packet
and the momentum transfer is purely transverse, i.e.
${\vec q}=(q_x,q_y,0)$. Furthermore, we chose a wave packet
that is a plane wave (or very delocalized) 
in the $z$-direction, i.e. $p_z$ of the wave packet is very
sharply peaked around $P_z$, and we choose $P_z$ such that
$P_z \gg M, |{\vec q}|$.
Then the abovementioned corrections due to the wave packet
can be made small without leading to large
relativistic corrections, which are governed by the expansion parameter 
$\frac{ {\vec q}\cdot {\vec p}}{E_{\vec p}^2}
\sim \frac{q_\perp}{L_\perp P^2}$.

In other words, if we consider a wave packet which is localized in the
transverse direction only, but a plane wave with very large momentum
in the $z$ direction, then as long as this system is probed with only
a transverse momentum transfer, the relativistic corrections to the
form factor of this wave packet are governed not by the Compton wavelength
but rather by $\lambda_P \equiv 1/\sqrt{m^2+p_z^2}$, which can be made
arbitrarily small.\footnote{Note that this result is reminiscent of the
result that in the infinite momentum frame, for purely transverse
momentum transfer, only terms that are diagonal in Fock space contribute
to the matrix elements of the (`good') current \cite{brodsky}.}
One thus finds for purely transverse momentum transfers in the
IMF
\be
{\cal F}_\Psi ({\vec q})=F({\vec q}^2).
\label{eq:formimf}
\ee

Physically, this implies that in the IMF
one can identify the (Fourier transform of)
the charge distribution in the transverse direction 
(the `transverse profile') with the formfactor
without relativistic corrections.
Of course, for ordinary form factors, this result is not very important since
the IMF is not a natural frame for physical
interpretation of the form factor. However, the analogous result will be
crucial when
we analyze (off forward) parton distributions for which the natural frame
{\it is} the IMF.

\section{Off Forward Parton Distributions}
Consider a wave packet $|\Psi \rangle$ which is chosen such 
that it has a sharp longitudinal momentum $p_z$, 
but whose position is a localized  wave packet in the 
transverse direction
\be
\left| \Psi \right\rangle = \int \frac{d^2p_\perp}
{\sqrt{2E_{\vec p}(2\pi)^2}} \Psi({\vec p}_\perp) \left|
{\vec p}\right\rangle
\label{eq:Psi}
\ee
Clearly,
\be
f_\Psi(x,{\vec b}_\perp) \equiv 
\int \frac{dx^-}{4\pi} 
\langle \Psi | \bar{\psi}({\vec b}_\perp,0) \gamma^+
\psi({\vec b}_\perp, x^-) |\Psi\rangle e^{ixp^+x^-} , 
\label{eq:fpsi}
\ee
describes the probability to find partons with momentum 
fraction $x$ at transverse (position) coordinate 
${\vec x}_\perp$ in this wave packet.
What we will show in the following is that 
$f_\Psi (x,{\vec b}_\perp)$
can be related to off-forward parton distribution functions
with $\zeta =0$.

Using Eq. (\ref{eq:Psi}), one finds
\bea
{\cal F}_\Psi (x,{\vec q}_\perp)&\equiv& \int d^2q_\perp 
e^{-i{\vec q}_\perp \cdot {\vec b}_\perp} 
f_\Psi (x,{\vec b}_\perp) \nonumber\\
&=& \int \frac{d^2p_\perp
\Psi^*({\vec p}^\prime_\perp) \Psi({\vec p}_\perp)}
{\sqrt{2E_{\vec p} 2 
E_{{\vec p}^\prime}}} \times \nonumber \\
& &\int dx^- e^{ixp^+x^-} 
\langle p^\prime | 
\bar{\psi}(0,{\vec 0}_\perp) \psi (x^-,{\vec 0}_\perp) |p\rangle
\nonumber\\
&=& \int \frac{d^2p_\perp
\Psi^*({\vec p}^\prime_\perp) \Psi({\vec p}_\perp)
}{\sqrt{2E_{\vec p} 2 
E_{{\vec p}^\prime}}}
f_\zeta(x, q^2).
\label{eq:fourier}
\eea
where ${\vec p}_\perp^\prime = {\vec p}_\perp +
{\vec q}_\perp$ and $p_z^\prime = p_z$, i.e. $\zeta=0$. 

\subsection{Nonrelativistic Limit}
Again we start by investigating the nonrelativistic limit
first, where one finds 
$E_{\vec p}\approx E_{{\vec p}^\prime}\approx m$ and therefore also
$q^2\approx -{\vec q}^2$. As a result, Eq. (\ref{eq:fourier}) simplifies,
yielding
\be
{\cal F}_\Psi (x,{\vec q}_\perp)= 
f(x, -{\vec q}_\perp^2)\int \frac{d^2p_\perp
\Psi^*({\vec p}^\prime_\perp) \Psi({\vec p}_\perp)
}{2m} .
\label{eq:fouriernr}
\ee
In order to proceed further, we choose a wave-packet that
is very localized in transverse position space.
Specifically, we choose a packet whose width in 
transverse momentum space is much larger than a typical 
QCD scale. That way, the dependence of the integrand
in Eq. (\ref{eq:fouriernr}) on ${\vec q}_\perp$ is mostly
due to the matrix element and not due to the wave packet $\Psi $.
Therefore, by making the wave packet very localized in position
space one obtains
\be
{\cal F}_\Psi (x,{\vec q}_\perp) =f(x, -{\vec q}_\perp^2),
\ee
and, just as it was the case for the form factor, it is
thus legitimate to identify the Fourier transform of
the $\zeta=0$ OFPD with respect to ${\vec q}_\perp$ with the
impact parameter dependence of the parton distribution in a very localized
localized wave packet, i.e. with the with the 
impact parameter dependence of the parton distribution in the target 
particle itself.

\subsection{Infinite Momentum Frame}

In an arbitrary frame, e.g. the rest frame,
relativistic corrections also spoil the
above identification of (Fourier transforms of) the impact parameter
dependence of parton distributions with OFPD at $\zeta=0$.
Similarly to the relativistic corrections for form factors, the above
identification becomes ambiguous when one looks at scales smaller than
the Compton wavelength of the target.

However, since the natural frame to think about (off forward) parton 
distributions is the IMF, we will skip details about
relativistic corrections in the rest frame and proceed immediately to
the IMF.
The crucial steps are as follows: 
\begin{itemize}
\item In Eq. (\ref{eq:fourier}), we choose a wave packet 
$\Psi({\vec p}_\perp)$ whose typical momentum scale $\lambda_\Psi$
is much smaller
than $\sqrt{m^2+p_z^2}$ yet much larger than the expected ${\vec q}^2$
dependence of $f(x,-{\vec q}^2)$, which should be on the order
of $\Lambda_{QCD}$.
\item we consider only momentum transfers that are smaller than 
$\lambda_\Psi$, i.e. we only probe the target with 
\end{itemize}
Of course, satisfying these requirements simultaneously is only possible
for $p_z \gg m$.

For a wave packet satisfying the above requirements, it is clear that
one can approximate $E_{\vec p} \approx E_{{\vec p}^\prime} 
\approx |p_z| $, as well as $q^2=-{\vec q}^2$ in Eq. (\ref{eq:fourier}), 
yielding
\bea
{\cal F}_\Psi (x,{\vec q}_\perp)&=& 
f(x, -{\vec q}_\perp^2)
\frac{1}{2|p_z|}
\int d^2p_\perp
\Psi^*({\vec p}^\prime_\perp) \Psi({\vec p}_\perp)
\nonumber\\
&=& f(x, -{\vec q}_\perp^2) \frac{1}{2|p_z|} ,
\label{eq:fourierimf}
\eea
where in the last step we used the fact that we had chosen
a very localized wave packet, i.e. 
\be\int d^2p_\perp \Psi^*({\vec p}_\perp+{\vec q}_\perp) \Psi({\vec p}_\perp)
\approx 
\int d^2p_\perp \Psi^*({\vec p}_\perp) \Psi({\vec p}_\perp)=1\ee
for ${\vec q}_\perp^2 = {\cal O}\left(\Lambda_{QCD}^2\right)$.
In the previous section we had argued that Eqs. (\ref{eq:formnr}) and 
(\ref{eq:formimf}) justify to identify the elastic form factor
$F({\vec q}^2)$ with the Fourier transform of the
charge distribution in the rest frame (nonrelativistic)
and the transverse charge distribution in the infinite
momentum frame respectively. In the same vein, 
Eqs. (\ref{eq:fouriernr}) and (\ref{eq:fourierimf})
justify to identify $f_{\zeta=0}(x,t)$ with the 
Fourier transform of (impact parameter dependent)
parton distribution functions with respect to the impact
parameter.

Note that, while it would seem unnatural to identify the elastic form
factor with something defined in the IMF, the natural frame
to think about parton distribution functions (forward
and off-forward) {\sl is} the IMF. Therefore, the fact that
Eq. (\ref{eq:fourierimf}) is free of relativistic corrections
only in the IMF, does not represent a serious restriction at all.

\section{$Q^2$ Evolution}
Throughout this paper we have suppressed the dependence of the
parton distributions on the momentum scale $Q^2$. Obviously, because of
scaling violations, all parton distributions involved depend
on $Q^2$ as well, e.g. $f(x,{\vec b}_\perp)$ should be replaced
by $f(x,{\vec b}_\perp, Q^2)$ and $f(x,-{\vec q}_\perp^2)$ should be replaced
by $f(x,-{\vec q}^2_\perp, Q^2)$.

Fortunately, it is rather straightforward to generalize our results
to take $Q^2$ evolution into account since the $Q^2$ evolution Eqs. for 
OFPDs reduce to the usual DGLAP
equations equations for $\zeta=0$ \cite{ji,ar}.
Of course, although all parton distributions that enter the DGLAP
equations for OFPDs depend on the invariant momentum transfer $t$, 
the evolution equations themselves are impact parameter
independent. 

Likewise, the impact parameter
dependent parton distributions evolve according to the
standard DGLAP equations es well in the sense that the same DGLAP
equation applies to each ${\vec b}_\perp$ and different ${\vec b}_\perp$ do
not mix under DGLAP evolution. To see this, one can use translational
invariance to shift the ${\vec b}_\perp)$-dependence on the r.h.s. of
Eq. (\ref{eq:fpsi}) from the operator to the state, i.e. instead of
measuring the correlator $\bar{\psi}({\vec b}_\perp,0)\gamma^+
\psi({\vec b}_\perp, x^-)$ in a wave packet centered around ${\vec 0}_\perp$ 
one can equivalently measure the
correlator $\bar{\psi}({\vec 0}_\perp,0)\gamma^+
\psi({\vec 0}_\perp, x^-)$ in a wave packet centered around $-{\vec b}_\perp$.

Combining these observations it is thus trivial to see that the 
identification of impact parameter dependent parton distributions with Fourier
transforms of $f(x,-{\vec q}^2)$ w.r.t ${\vec q}$ is preserved
under QCD evolution in the sense that
\be
f(x_{Bj}, {\vec b_\perp},Q^2) = \int \frac{d^2q_\perp}{2\pi}
e^{i{\vec q}_\perp {\vec b}_\perp}
f_{\zeta=0}(x_{Bj}, -{\vec q}_\perp^2,Q^2) 
\ee
is valid for all $Q^2$ (as long as $Q^2$ is large enough for DGLAP to be
applicable).

\section{Summary and Discussion}
Off-forward parton distributions at $\zeta=0$ allow a 
simultaneous measurement of the light-cone momentum and transverse
position (impact parameter!) distribution of partons in a hadron:
\be
f(x_{Bj}, {\vec b_\perp}) = \int \frac{d^2q_\perp}{2\pi}
e^{i{\vec q}_\perp {\vec b}_\perp}
f_{\zeta=0}(x_{Bj}, -{\vec q}_\perp^2) .
\label{eq:final}
\ee
This fundamental observation is strictly true in the
IMF, but receives relativistic corrections
in other frames. Those corrections are of the same nature as
the relativistic corrections that spoil the identification of
the charge form factor with the Fourier transform of a charge
distribution for systems where the Compton wavelength
is of the same order as the size, i.e. $MR={\cal O}(1)$, or
larger. Of course in nonrelativistic systems, the identification
of $f_{\zeta=0}(x_{Bj}, {\vec q}_\perp^2)$ with the Fourier
transform of the longitudinal momentum/transverse position
distribution function is also strictly true.

Moreover, although we restricted our discussion of spin independent
parton distribution functions, it should be clear that our result
generalizes to spin dependent distribution as well.

While these result is not so much of importance for exact
calculations of off-forward distribution functions (for 
example within the framework of lattice QCD),
the main application of our result
lies both more within the areas modeling, phenomenology
as well as the physical interpretation of experimental
and numerical (lattice) data.

However, the most important application is using experimental 
(or numerical) data on the $t$ dependence to learn how parton 
distributions depend on the impact parameter.
For example, by considering the slope of $f_\zeta(x,t)$ 
w.r.t. $t$ at $t=\zeta=0$ one obtains the parton distribution
weighted by the impact parameter squared
and thus the `outer' region of the target hadron
gets more strongly emphasized. A precise
measurement of this slope could thus reveal important 
information in the transverse distribution of partons within
hadrons, which could also help to distinguish surface effects
from bulk effects in nucleons and nuclei.

More specific applications, should also include extending models for
conventional parton distribution functions to off-forward
distributions at $\zeta \rightarrow 0$. However, providing
explicit examples for this is beyond the scope of this paper.

\noindent {\bf Acknowledgments:}
It is a pleasure to thank R.L. Jaffe, X. Ji, A.R. Radyushkin, and
A.I. Vainshtein for helpful and encouraging discussions. 
This work was supported in part by a grant from
DOE (DE-FG03-95ER40965) and in part by TJNAF.


\begin{references}
\bibitem{ji} X. Ji, Phys.\ Rev.\ Lett.\ {\bf 78}, 610 (1997);
 Phys. Rev. D\ {\bf 52}, 3841 (1995).
\bibitem{ar} A.V. Radyushkin, Phys.\ Rev.\ D\ {\bf 56}, 5524 (1997).
\bibitem{wally} X. Ji, W. Melnitchouk, and X. Song,
Phys.\ Rev.\ D\ {\bf 56}, 5511 (1997).
\bibitem{kroll1} M. Diehl et. al, Phys.\ Lett.\ B\ {\bf 460}, 204 (1999). 
\bibitem{kroll2} M. Diehl et. al, Eur.\ Phys.\ J.\ {\bf C8}, 409 (1999) 
\bibitem{ar3} A.V. Radyushkin, Phys.\ Rev.\ D\ {\bf 58}, 114008 (1998).
\bibitem{ar2} A.V. Radyushkin, Proc. of the Workshop
``Physics and Instrumentation with 6-12 GeV Beams'',
Jefferson Lab, June 15-18, 1998.
\bibitem{brodsky} S.J. Brodksy in `QCD, Lightcone Physics and
Hadron Phenomenology', Eds. C.R. Ji and D.-P. Min,
World Scientific, Singapore, 1998.
\end{references}
\end{document}